\documentclass[12pt]{article}
\begin{document}
\baselineskip=24pt
\bibliographystyle{plain}
\newcommand{\subs}[1]{\ensuremath{\scriptstyle \mathit{#1}}}
\newcommand{\comm}[1]{\ensuremath{\mbox{\textup{#1}}}}

\title{Parity-Preserving Pauli-Villars Regularization in 
       $(2+1)$-Dimensional Gauge Models\footnote{
       Physics of Atomic Nuclei, Vol. 58, No. 9, 1995, pp. 1619-1621.
       Translated from Yadernaya Fizika, Vol. 58, No. 9, 1995, pp. 1718-1720.}}

\bigskip

\bigskip

\author{
L. R. Baboukhadia, A. A. Khelashvili, and N. A. Kiknadze \\ 
Tbilisi State University \\ Tbilisi 380028 \\  Georgia}
\date{}
\maketitle
\bigskip

\bigskip

\begin{abstract}
\baselineskip=24pt

\noindent It is shown that, in QED$_{3}$, the Pauli-Villars regularization 
involving a pair of auxiliary fermion fields with masses of opposite sign 
leads to results that are consistent with those obtained using all other 
parity-preserving schemes of regularization. At the same time, ambiguity 
problems remain unsolved in non-Abelian models (QCD$_{3}$).
\end{abstract}

\vfill


\vfill

\newpage




In the past few years, the problem of the ambiguity of various regularization
schemes in $(2+1)$-dimensional gauge models has become the subject of lively 
discussions. This problem dates back to the study of Deser {\em et al}. [1], 
who showed that, at the one-loop level of perturbation theory, 
Pauli-Villars (PV) 
regularization and dimensional regularization lead to radically different 
physical patterns in QED$_{3}$ and QCD$_{3}$. Further investigations based 
on other schemes of regularization (such as analytic regularization [2], 
dimensional regularization [3,4], proper-time regularization [5], etc.), as 
well as new analyses in the PV scheme [6,7], confirmed this discrepancy: in 
the PV scheme, the initially massless gauge boson remains massless in both 
Abelian and non-Abelian theories with a massive fermion, while in all other 
regularization schemes, this is not the case. In theories with a massless 
fermion, there arises a discrepancy in the Chern-Simons (CS) term.

\smallskip

It was emphasized in [6] that the reason behind this discrepancy is that PV 
regularization violates parity, while other schemes of renormalization 
preserve parity.

\smallskip

It is obvious that preference cannot be given to any scheme of regularization
(and, hence, to any specific quantum theory) without invoking some additional
criteria. Only for the theory with a massless fermion can we draw conclusions 
with a certain degree of confidence. In the absence of the CS term, the 
classical Lagrangian density of this theory is invariant under all discrete
($P$, $T$) symmetries. In the one-loop approximation, neither the gauge-boson
propagator, nor the three-gluon vertex involves sources of the antisymmetric
tensor $\varepsilon_{\subs{\mu\nu\lambda}}$ (or the CS term) prior to 
applying a regularization procedure. Naturally, the emergence of such terms is
an artifact of the PV scheme.

\smallskip

Here, we would like to take a fresh look at this ambiguity. In the Abelian 
case, we construct a nonminimal version of PV regularization. This procedure 
leads to a result that is consistent with the results of other regularization 
schemes. In the non-Abelian case, however, our method violates gauge symmetry.

\smallskip

We consider the $(2+1)$-dimensional gauge theory with a massless fermion. The 
Lagrangian density of the theory is given by
\begin{equation}
  \label{eq:lag}
    L = - \frac{1}{4} F_{\subs{\mu\nu}}F^{\subs{\mu\nu}} + 
        \overline{\Psi}(i\hat{\partial} + g\hat{A})\Psi \, .
\end{equation}

\smallskip

The photon polarization operator has the form
\begin{equation}
  \label{eq:pol}
    \Pi_{\subs{\mu\nu}}(k) = 
      \left( g_{\subs{\mu\nu}} \, - \, 
        \frac{k_{\mu}k_{\nu}}{k^{2}} \right) \Pi^{(1)}(k^{2}) +
      i\varepsilon_{\subs{\mu \nu \lambda}}k^{\lambda}\Pi^{(2)}(k^{2}) \, . 
\end{equation}

\smallskip

The structure functions $\Pi^{(1,2)}(k^{2})$ correspond 
to parity-preserving and parity-violating contributions, respectively, and 
determine the photon propagator [1]
\begin{equation}
  \label{eq:ddd}
    D_{\subs{\mu\nu}}(k) = 
      \frac{-i}{k^{2} - \Pi(k^{2})}
      \left[  
	g_{\subs{\mu\nu}} - \frac{k_{\mu}k_{\nu}}{k^{2}} -
	i\varepsilon_{\subs{\mu \nu \lambda}}\frac{k^{\lambda}}{k^{2}}
          \mathcal{M}(k^{2})
      \right],
\end{equation}
where
\begin{equation}
  \label{eq:ppp}
    \Pi(k^{2}) = \Pi^{(1)}(k^{2}) +
                 \left[ \Pi^{(2)}(k^{2}) \right]^{2}
                 \left( 1 - \frac{\Pi^{(1)}(k^{2})}{k^{2}} \right)^{-1} \, ,
\end{equation}

\smallskip

\begin{equation}
  \label{eq:mmm}
    \mathcal{M}(k^{2}) = \Pi^{(2)}(k^{2})
		 \left( 1 - \frac{\Pi^{(1)}(k^{2})}{k^{2}} \right)^{-1} \, .
\end{equation}

\smallskip

The model is superrenormalizable; that is, divergencies arise only at the 
one-loop level. In the one-loop approximation, we have

\begin{equation}
  \label{eq:int}
    \Pi_{\subs{\mu\nu}}(k) = -ig^{2}
      \int \frac{d^{3}q}{(2\pi)^{3}} 
	\comm{tr}\left[
	     \gamma_{\mu} S\left( k+q \right) \gamma_{\nu} S \left( q \right)
	  \right] \, ,
\end{equation}
where $S\left(q\right) = i\hat{q}^{\subs{-1}}$ to the same degree of 
accuracy. It is obvious that $\Pi_{\subs{\mu\nu}}\left( k \right)$ diverges
linearly.

\smallskip

In the non-Abelian case, we also have to consider the three-gluon vertex

\begin{equation}
  \label{eq:vertex}
    \Pi_{\subs{\mu\nu\lambda}}(q,k) = -g^{3}
      \int \frac{d^{3}p}{( 2\pi )^{3}} \comm{tr} 
        \left[
	  \gamma_{\mu} S(p) \gamma_{\nu} S(p+k)  \gamma_{\lambda}
	  S(p+k+q)
	\right] \, ,
\end{equation}
which diverges logarithmically.

Deser {\em et al} [1], who discussed the theory with a massive fermion, 
showed that, in the one-loop approximation, the factor $\Pi^{(2)}(k^{2})$
(which does not diverge) suffers from the ambiguity of regularization: PV
regularization yields $\Pi^{(2)}(0) = 0$, while dimensional regularization
leads to a nonzero value of $\Pi^{(2)}(0)$. As a CS term arises in this 
theory, the ambiguity of regularization has a bearing on problems that are
of prime physical importance.

\smallskip

In our opinion, an important point was highlighted in the study of Shifman [8],
who defined the massless theory as the zero-mass limit of the theory with a
massive fermion, since the two-fermion phase-space volume involves an infrared
singularity. He showed that this limiting transition is not smooth; that is, 
the residual term in the odd form factor $\Pi^{(2)}(k^{2})$ depends on the
sign of the fermion mass [1] just as it depends on the sign of the 
auxiliary-fermion mass when PV regularization is used.

\smallskip

To shed new light on this problem and to cover both cases, we consider the
theory with a massive fermion. To do this, we replace the propagator $S(q)$ by
$S(q,m) = i(\hat{q}-m)^{-1}$. Following integration with respect to momenta in
Euclidean metric with the cutoff parameter $\Lambda$, the polarization 
operator assumes the form [1,5]
 
\begin{equation}
  \label{eq:massive}
    \Pi_{\subs{\mu\nu}}(k, m) = - \frac{g^{2}}{3\pi} \Lambda g_{\subs{\mu\nu}}
      + \left( g_{\subs{\mu\nu}} \, - \, \frac{k_{\mu}k_{\nu}}{k^{2}} \right)
        \widetilde{\Pi}^{(1)}(k^{2},m) +
      i\varepsilon_{\subs{\mu\nu\lambda}}k^{\lambda}
       \widetilde{\Pi}^{(2)}(k^{2},m) \, , 
\end{equation}
where
\begin{eqnarray}
  \label{eq:massive1}
    \lefteqn{ \nonumber
    \widetilde{\Pi}^{(1)}(k^{2},m) = - \frac{g^{2}}{2\pi}
      \int_{0}^{1} d\alpha \frac{\alpha(1-\alpha)k^{2}}
				{\sqrt{m^{2}-k^{2}\alpha(1-\alpha)}} =
     } \\
     & &
      - \frac{g^{2}}{2\pi} \left[
	\frac{\sqrt{m^{2}}}{2} - \left(\frac{m^{2}}{2}+\frac{k^{2}}{8}\right)
          \frac{1}{\sqrt{k^{2}}} \ln{\left( \frac{2\sqrt{m^{2}}+\sqrt{k^{2}}}
					        {2\sqrt{m^{2}}-\sqrt{k^{2}}}
				    \right)}
			   \right] \, ,
\end{eqnarray}

\smallskip

\begin{equation}
  \label{eq:massive2}
    \widetilde{\Pi}^{(2)}(k^{2},m) = - \frac{g^{2}}{4\pi}
      \int_{0}^{1} d\alpha \frac{m}
				{\sqrt{m^{2}-k^{2}\alpha(1-\alpha)}} =
      - \frac{g^{2}}{4\pi} \frac{m}{\sqrt{k^{2}}} 
	\ln{\left( \frac{ 2\sqrt{m^{2}}+\sqrt{k^{2}} }
		        { 2\sqrt{m^{2}}-\sqrt{k^{2}} }
	    \right)} \, .
\end{equation}

\smallskip

In passing, we note that divergencies do not arise in dimensional 
regularization; hence, the result is given by expressions (9) and (10). In the
case of PV regularization, it is necessary to go over to the limit 
$|m| \rightarrow \infty$ in expressions (9) and (10). We have

\begin{equation}
  \label{eq:limit1}
    \widetilde{\Pi}^{(1)}( k^{2}, |m|\rightarrow \infty ) = 
       - \frac{g^{2}}{12\pi}\frac{k^{2}}{\sqrt{m^{2}}} \longrightarrow 0 \, ,
\end{equation}

\smallskip

\begin{equation}
  \label{eq:limit2}
    \widetilde{\Pi}^{(2)}( k^{2}, |m|\rightarrow \infty ) = 
       \frac{g^{2}}{4\pi} \frac{m}{\sqrt{m^{2}}} = 
       \frac{g^{2}}{4\pi} \comm{sgn}(m) \neq 0 \, .
\end{equation}

\smallskip 

Therefore, the PV regularization using one mass $M$ yields

\begin{equation}
  \label{eq:pv1}
    \Pi^{(1)}_{Reg}(k^{2},m) =  \widetilde{\Pi}^{(1)}(k^{2},m) \, ,
\end{equation}

\smallskip

\begin{equation}
  \label{eq:pv2}
    \Pi^{(2)}_{Reg}( k^{2}, m) =  \widetilde{\Pi}^{(2)}( k^{2}, m ) -
				   \frac{g^{2}}{4\pi} \comm{sgn}(M) \, .
\end{equation}

\smallskip

For the massless theory, these expressions become:

\begin{equation}
  \label{eq:massless1}
    \Pi^{(1)}_{Reg}( k^{2}) = \Pi^{(1)}(k^{2}, m=0) = 
			      \frac{g^{2}\sqrt{-k^{2}}}{16} \, ,
\end{equation}

\smallskip

\begin{equation}
  \label{eq:massless2}
    \Pi^{(2)}_{Reg}( k^{2}) = \Pi^{(2)}( k^{2}, m = 0 ) = 
                              - \frac{g^{2}}{4\pi} \comm{sgn}(M) \, .
\end{equation}

\smallskip

At the same time, dimensional regularization leads to the expressions

\begin{equation}
  \label{eq:dimreg}
    \Pi^{(1)}(k^{2}) = \frac{g^{2}\sqrt{-k^{2}}}{16} \, ,
    \;\;\;\;\;\;\;\;\;\;\;
    \Pi^{(2)}(k^{2}) = 0 \, .
\end{equation}

\smallskip

Thus, in the theory with a massless fermion, dimensional regularization does
not generate an odd structure, that is $\Pi^{(2)} = 0$. In the PV method, we 
have $\Pi^{(2)} \neq 0$, and the CS term arises. However, the gauge-field
propagator does not develop a nontrivial pole as a result, because relations
(4) and (5) yield the values $\Pi(0) = 0$ and $\mathcal{M}(0) = 0$ for
$\Pi^{(1)} \sim \sqrt{-k^{2}}$; that is, the mass spectra of the gauge field
coincide in all methods of regularization. In other words, the dynamical 
violation of parity in the PV method does not lead to the dynamical 
generation of the gauge-field mass. At the same time, the spectra differ in the
theory with a massive fermion [1].

\smallskip

The question now arises of whether it is possible to modify the PV scheme in 
such a way as to get rid of the spontaneous generation of $\Pi^{(2)}$. For
this purpose, we consider a larger number of auxiliary regularizing fermions;
that is, we define the regularized propagator as 

\begin{equation}
  \label{eq:reg}
    \Pi^{R}_{\subs{\mu\nu}}(k) = 
	\sum_{i=1}^{N} c_{i} \Pi_{\subs{\mu\nu}}(k,M_{i}) \, ,
\end{equation}
where $c_{1} = 1$, $M_{1} = m$, and $|M_{i}| \rightarrow \infty$ for $i\geq2$.

\smallskip

The divergence in (8) can be eliminated by imposing the condition
\begin{equation}
  \label{eq:cond}
     \sum_{i=1}^{N} c_{i} = 1 + \sum_{i=2}^{N} c_{i} = 0 \, .
\end{equation}

\smallskip

To ensure the absence of an antisymmetric structure, we must additionally 
require fulfillment of the equality

\begin{equation}
  \label{eq:cond1}
     \sum_{i=2}^{N} c_{i} \comm{sgn}(M_{i}) = 0 \, .
\end{equation}

\smallskip

If all auxiliary masses have the same sign, conditions (19) and (20) cannot
be satisfied simulteneously.

\smallskip

In order that these two equalities be consistent, the mass of at least one 
auxiliary fermion must differ in sign from the masses of other auxiliary 
fermions. It can easily be shown that, in such cases, the coefficients $c_{i}$
are not integral. For example, in the case of two masses satisfying the 
condition $M_{2} = -M_{3} = M$, equalities (19) and (20) yield

\begin{equation}
  \label{eq:coeff}
     c_{2} = c_{3} = -1/2 \, .
\end{equation}

\smallskip

It is difficult to interpret nonintegral coefficients if it is necessary to
substantiate regularization with the aid of counterterms. For the counterterms
in the polarization operator, we can take the standard Lagrangian densities of
the massive fermion, setting the coupling constants in these Lagrangian 
densities to $|c_{i}|^{1/2}g$, rather than to $g$. This leads to the required
cancellation of the induced odd structure in the terms of order $g^{2}$. In
higher orders of perturbation theory, the contributions of the counterterms
themselves vanish in the limit $|M| \rightarrow \infty$. Therefore, in the
Abelian case, we can achieve the complete cancellation of the induced CS term
in the massless theory, the result that we strived to obtain.

\smallskip

In the non-Abelian case, the removal of regularization does not lead to the
complete disappearance of mass dependence in the three-gluon vertex (7). To be
more specific, we have [6]

\begin{equation}
  \label{eq:verfinal}
    \lim_{|M| \rightarrow \infty} \Pi_{\subs{\mu\nu\lambda}}(q,k;M) = 
	 -\varepsilon_{\subs{\mu \nu \lambda}} \frac{g^{3}}{4\pi}
	\comm{sgn}(M) \, .
\end{equation}

\smallskip

It is obvious that condition (21) ensures cancellation of this contribution;
however, it is impossible to match the counterterms required at this stage 
with those introduced previously because the auxiliary fermions must be 
assigned the charges $|c_{i}|^{1/3}g$. This would lead to violation of local
gauge invariance, as the coefficients $c_{i}$ are not integral.

\smallskip

Thus, including two auxiliary fermions with masses of opposite sign (or one
auxiliary Dirac fermion) in the regularization scheme, we can match the PV
regularization in Abelian electrodynamics (both in the massless and massive 
cases) with other schemes of regularization. An important point in this
analysis is the restoration of discrete symmetries $(P,T)$ at the 
intermediate stage.

\section*{Acknowledgements}

We are grateful to G. Japaridze and K. Turashvili for stimulating discussions.

\eject

\end{document}